\pdfoutput=1    
\documentclass[%
superscriptaddress,
 amsmath,amssymb,
 aps,
 twocolumn
]{revtex4-1}

\usepackage{graphicx}
\usepackage{bm}
\usepackage{braket}
\usepackage{subfiles}
\usepackage{array}
\usepackage{float}
\usepackage{multirow}
\usepackage{amsmath}
\usepackage{amssymb}
\usepackage{graphicx}
\PassOptionsToPackage{normalem}{ulem}
\usepackage{ulem}
\usepackage{qcircuit}

\newcommand{\hShift}[0]{&\push{\rule{5.2em}{0em}}}
\newcommand{\inData}[0]{ \hShift&\lstick{Z^aX^b\ket{\phi}} }
\newcommand{\inAuxa}[1]{ \hShift&\lstick{Z^{#1}P^a\ket{+}} }
\newcommand{\inAuxb}[1]{ \hShift&\lstick{Z^{#1}P^b\ket{+}} }
\newcommand{\decZX} [2]{ \gate{Z^{#1}X^{#2}} }

\newcommand{\eqAuxHOM}{
    \begin{align*}
     & \frac{1}{2}\left(\hat{a}_{H}^{\dagger}+e^{i\alpha}\hat{a}_{V}^{\dagger}\right)\left(\hat{b}_{H}^{\dagger}+e^{i\beta}\hat{b}_{V}^{\dagger}\right)\\
    \to & \frac{1}{2\sqrt{2}}\hat{a}_{H}^{\dagger}\left(\hat{b}_{H}^{\dagger}+\hat{b}_{V}^{\dagger}\right)-\frac{e^{i\left(\alpha+\beta\right)}}{2\sqrt{2}}\hat{a}_{V}^{\dagger}\left(\hat{b}_{H}^{\dagger}-\hat{b}_{V}^{\dagger}\right)
    \end{align*}
}

\newcommand{\eqAuxPS}{
    \begin{align*}
    \text{Post-selecting on }\hat{b}_{V}^{\dagger}\text{: } & \left|\zeta\right>=\frac{1}{\sqrt{2}}\left(\left|0\right>+e^{i\left(\alpha+\beta\right)}\left|1\right>\right)\\
    \text{Post-selecting on }\hat{b}_{H}^{\dagger}\text{: } & \left|\zeta\right>=\frac{1}{\sqrt{2}}\left(\left|0\right>-e^{i\left(\alpha+\beta\right)}\left|1\right>\right)
    \end{align*}
}

\newcommand{\eqAuxNewKey}{
    \[
    a''=i+j+a\left(b+1\right)+\left(a+b\right)k_{2}+k_{1}
    \]
}
\newcommand{\figCliffordKeyTransforms}{
    \begin{table}
        \begin{centering}
        \begin{tabular}{|c|c|c|c|}
        \hline 
        \multicolumn{4}{|c|}{\textbf{Single Qubit Clifford Gates}}\tabularnewline
        \multicolumn{4}{|c|}{\Qcircuit @C=1em @R=0.7em{
        	&\gate{U}&\qw&\push{=\rule{0.6em}{0em}}&\gate{Z^a X^b}&\gate{U}&\gate{Z^{a'} X^{b'}}&\qw
        }}\tabularnewline
        \hline 
        \multicolumn{1}{|c}{\textbf{$\boldsymbol{U}$}} & \multicolumn{1}{c}{Matrix Representation} & \multicolumn{1}{c}{\textbf{$\boldsymbol{a'}$}} & \textbf{$\boldsymbol{b'}$}\tabularnewline
        \hline 
        $X$ & $\left(\begin{array}{cc}
        0 & 1\\
        1 & 0
        \end{array}\right)$ & \multirow{3}{*}{$a$} & \multirow{3}{*}{$b$}\tabularnewline
        \cline{1-2} 
        $Y$ & $\left(\begin{array}{cc}
        0 & -i\\
        i & 0
        \end{array}\right)$ &  & \tabularnewline
        \cline{1-2} 
        $Z$ & $\left(\begin{array}{cc}
        1 & 0\\
        0 & -1
        \end{array}\right)$ &  & \tabularnewline
        \hline 
        $H$ & $\frac{1}{\sqrt{2}}\left(\begin{array}{cc}
        1 & 1\\
        1 & -1
        \end{array}\right)$ & $b$ & $a$\tabularnewline
        \hline 
        $P$ & $\left(\begin{array}{cc}
        1 & 0\\
        0 & i
        \end{array}\right)$ & $a\oplus b$ & $b$\tabularnewline
        \hline 
        \hline 
        \multicolumn{4}{|c|}{\textbf{Two Qubit Clifford: CNOT}}\tabularnewline
        \multicolumn{4}{|c|}{\Qcircuit @C=1em @R=0.7em{
        	&\ctrl{1} &\qw&\push{=\rule{1.3em}{0em}}& \gate{Z^a X^b} & \ctrl{1} & \gate{Z^{a\oplus c} X^b} &\qw \\
        	&\targ	&\qw&\push{ \rule{2.4em}{0em}}& \gate{Z^c X^d} & \targ	& \gate{Z^c X^{b\oplus d}} &\qw
        }}\tabularnewline
        \hline 
        \end{tabular}
        \par\end{centering}
        \caption{Key transformation under Clifford gates. In circuit diagrams above, $Z^{a}X^{b}$ and $Z^{a'}X^{b'}$ are encrypting and decrypting maps performed by Alice.\label{tab:CliffordKeyTransform}}
    \end{table}
}

\newcommand{\figCanonicalCases}{
    \begin{table}
        \begin{centering}
        \begin{tabular}{|>{\raggedright}p{8cm}|}
        \hline 
        \tabularnewline
        \textbf{\uline{Case 1: \mbox{$\boldsymbol{b'=b}$}}}\uline{,
        use \mbox{$\left|\xi_{b}\right>=Z^{j}P^{b}\left|+\right>$}}\tabularnewline
        Example:\tabularnewline
        \Qcircuit @C=0.55em @R=0.55em{
        \inData	&\gate{T}	&\ctrl{1}	&\decZX{a''}{b''}	&\qw	&\rstick{T\ket{\phi}}\\ \inAuxb{j}	&\qw	&\targ	&\measureD{k}\cwx[-1]		
        }\tabularnewline
        $a''=j+b\left(k+1\right)+a$\\
        $b''=b$\tabularnewline
        \hline 
        \tabularnewline
        \textbf{\uline{Case 2: \mbox{$\boldsymbol{b'=a}$}}}\uline{,
        use \mbox{$\left|\xi_{a}\right>=Z^{i}P^{a}\left|+\right>$}}\tabularnewline
        Example:\tabularnewline
        \Qcircuit @C=0.55em @R=0.55em{
        	\inData&\gate{H}&\gate{T} &\ctrl{1} &\decZX{a''}{b''} &\qw &\rstick{TH\ket{\phi}} \\
        	\inAuxa{i}&\qw	 &\qw	  &\targ	&\measureD{k}\cwx[-1]
        }\tabularnewline
        $a''=i+a\left(k+1\right)+b$\\
        $b''=a$\tabularnewline
        \hline 
        \tabularnewline
        \textbf{\uline{Case 3: \mbox{$\boldsymbol{b'=a\oplus b}$}}}\uline{,
        use both \mbox{$\left|\xi_{a}\right>$} and \mbox{$\left|\xi_{b}\right>$}}\tabularnewline
        Example:\tabularnewline
        \Qcircuit @C=0.55em @R=0.55em{
        \inData	&\gate{P}	&\gate{H}	&\gate{T}	&\ctrl{1}	&\decZX{a''}{b''}	&\qw	&\rstick{THP\ket{\phi}}\\ \inAuxa{i}	&\qw	&\ctrl{1}	&\qw	&\targ	&\measureD{k_2}\cwx[-1]\\		 \inAuxb{j}	&\qw	&\targ	&\qw	&\qw	&\measureD{k_1}\cwx[-1]		
        }\tabularnewline
        $a''=i+j+a\left(b+1\right)+\left(a+b\right)k_{2}+bk_{1}$\\
        $b''=a+b$\tabularnewline
        \hline 
        \end{tabular}
        \par\end{centering}
        \caption{Table of ancilla qubit(s) usage. We denote by $a'$ and $b'$ secret keys $a$ and $b$, transformed by unitary preceding $T$-gate. Here, $i,j\in\left\{ 0,1\right\} $ are classical keys that secure $\left|\xi_{a}\right>$ and $\left|\xi_{b}\right>$ respectively. All arithmetic taken modulo 2.\label{tab:T-correction}}
    \end{table}
}

\newcommand{\qcspace}[1]{\push{\rule{#1}{0em}}}

\newcommand{\figTsimple}{
    \begin{figure}[h]
        \Qcircuit @C=1em @R=0.7em{
        	\qcspace{1em}&\gate{T}&\qw &\push{=\rule{0.5em}{0em}}&\gate{Z^{a'} X^{b'}} & \gate{T} & \gate{P^{b'}} & \gate{Z^{a'+b'} X^{b'}} & \qw
        }
        \caption{A non-Clifford gate, like $T$, potentially requires additional $\pi/2$ phase correction here written as a $P$ gate.\label{fig:TgatePcorrection}}
        
        \bigskip{}
        
        \Qcircuit @C=1em @R=0.7em{
        	\qcspace{6em}  &\lstick{Z^a X^b\ket{\phi}}  &\gate{U'_{\mathrm{Clifford}}}&\gate{T}  &\ctrl{1}&\rstick{Z^{a''}X^{b''}T\ket{\phi}}\qw  \\
        	\qcspace{6em}  &\lstick{Z^r P^s\ket{+}}     &\qw     &\qw       &\targ   &\measureD{k}  &\cw
        }
        \caption{Phase correction after a $T$ gate that does not require Alice to divulge encryption key $b$ to Bob. Decryption keys $a''$ and $b''$ now depend on $a$, $b$, $r$, $s$, $k$ and $U'_{\mathrm{Clifford}}$.\label{fig:figSemiSimpleT}}
    \end{figure}
}

\newcommand{\figComparatorBasic}{
    \begin{figure}
    \begin{centering}
    \Qcircuit @C=1em @R=0.7em{
    	\qcspace{9em}	&\lstick{\rho_\alpha}	&\ctrl{1}	&\gate{H}	&\meter\\
    	\qcspace{9em}	&\lstick{\rho_\beta}	&\targ	&\qw	&\meter
    }
    \par\end{centering}
    \caption{Simple comparator circuit\label{fig:Comparator}}
    \end{figure}
}

\newcommand{\figComparatorSecure}{
    \begin{figure}
    \begin{centering}
    \includegraphics[width=5cm]{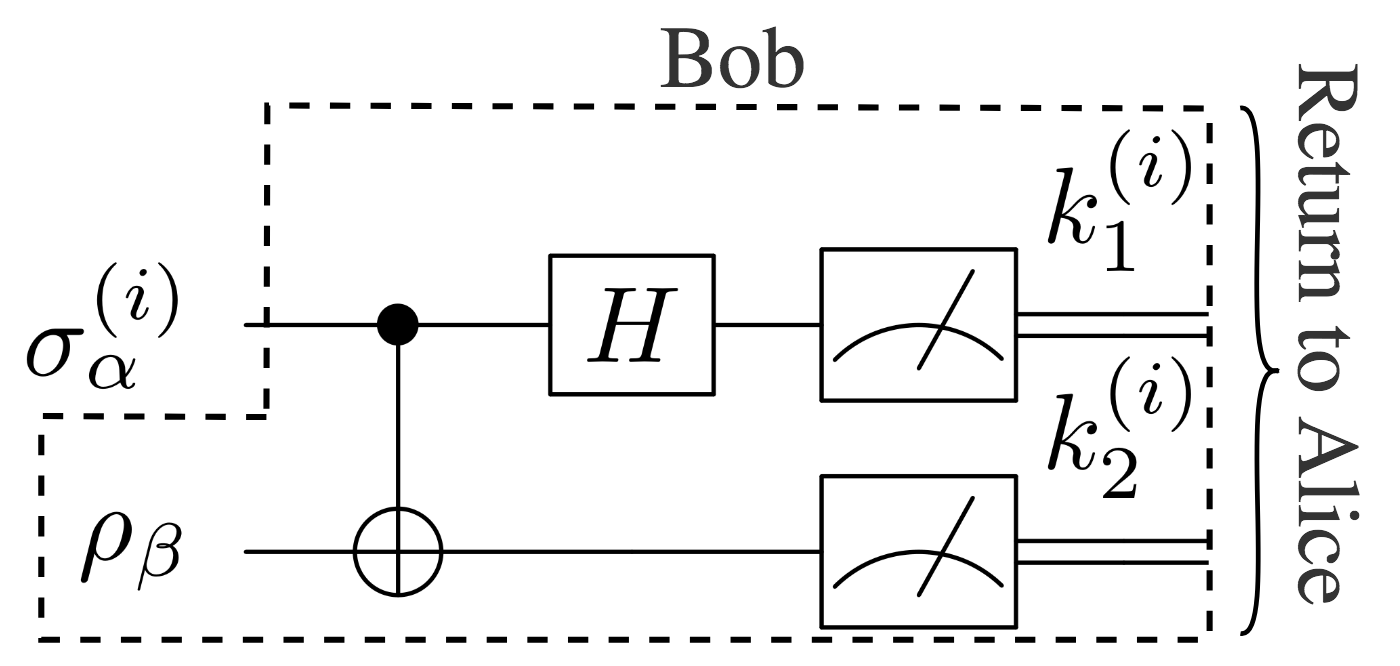}
    \par\end{centering}
    \caption{Secure comparator. Here, $\sigma_{\alpha}^{\left(i\right)}$ is the
    $i$-th encrypted copy of Alice's qubit, $\rho_{\beta}$ is Bob's
    qubit, and $k_{1}^{\left(i\right)},k_{2}^{\left(i\right)}\in\left\{ 0,1\right\} $
    are classical measurement outcomes.\label{fig:Secure2PartyComparator}}
    \end{figure}
}

\newcommand{\figExperimentSchematic}{
    \begin{figure}
        \begin{centering}
        \includegraphics[width=8cm]{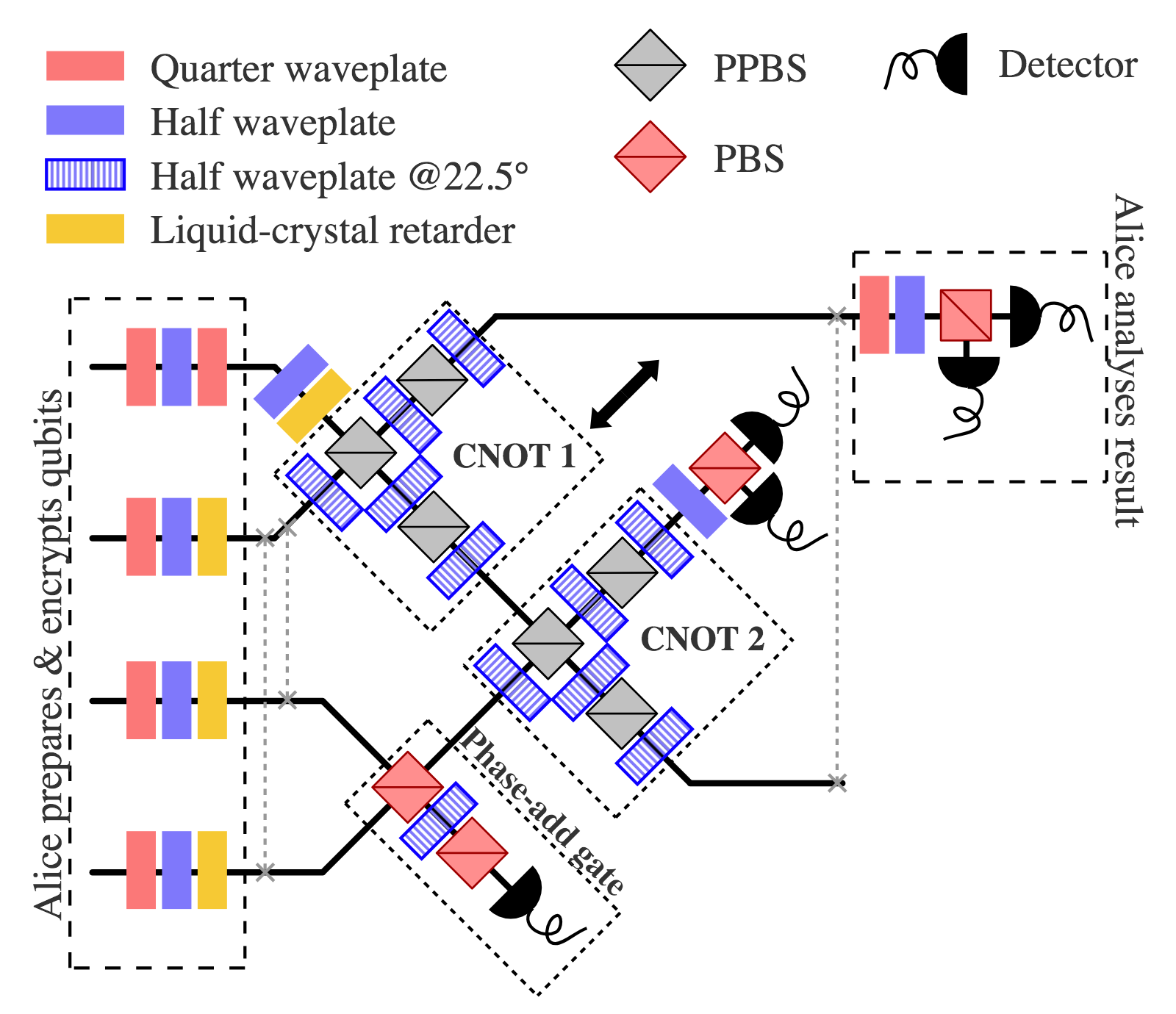}
        \par\end{centering}
        \caption{Schematic of optical circuit designed to implement each canonical
        case enumerated in Table \ref{tab:T-correction}. Each incoming rail
        from the left is a separate photon from an SPDC event (see Figure
        \ref{fig:Source-schematic} for source schematic). Liquid-crystal
        retarders allow us to modulate the phase of a qubit much more quickly
        and precisely than a motorised waveplate mount. Two-qubit gates shown
        here in dashed boxes can be bypassed as necessary, either by swapping
        a photon onto another rail or by translating ``CNOT 1'' out of the
        optical path.\label{fig:Experimental-setup}}
    \end{figure}
}

\newcommand{\figSource}{
    \begin{figure}
        \centering{}\includegraphics[width=8cm]{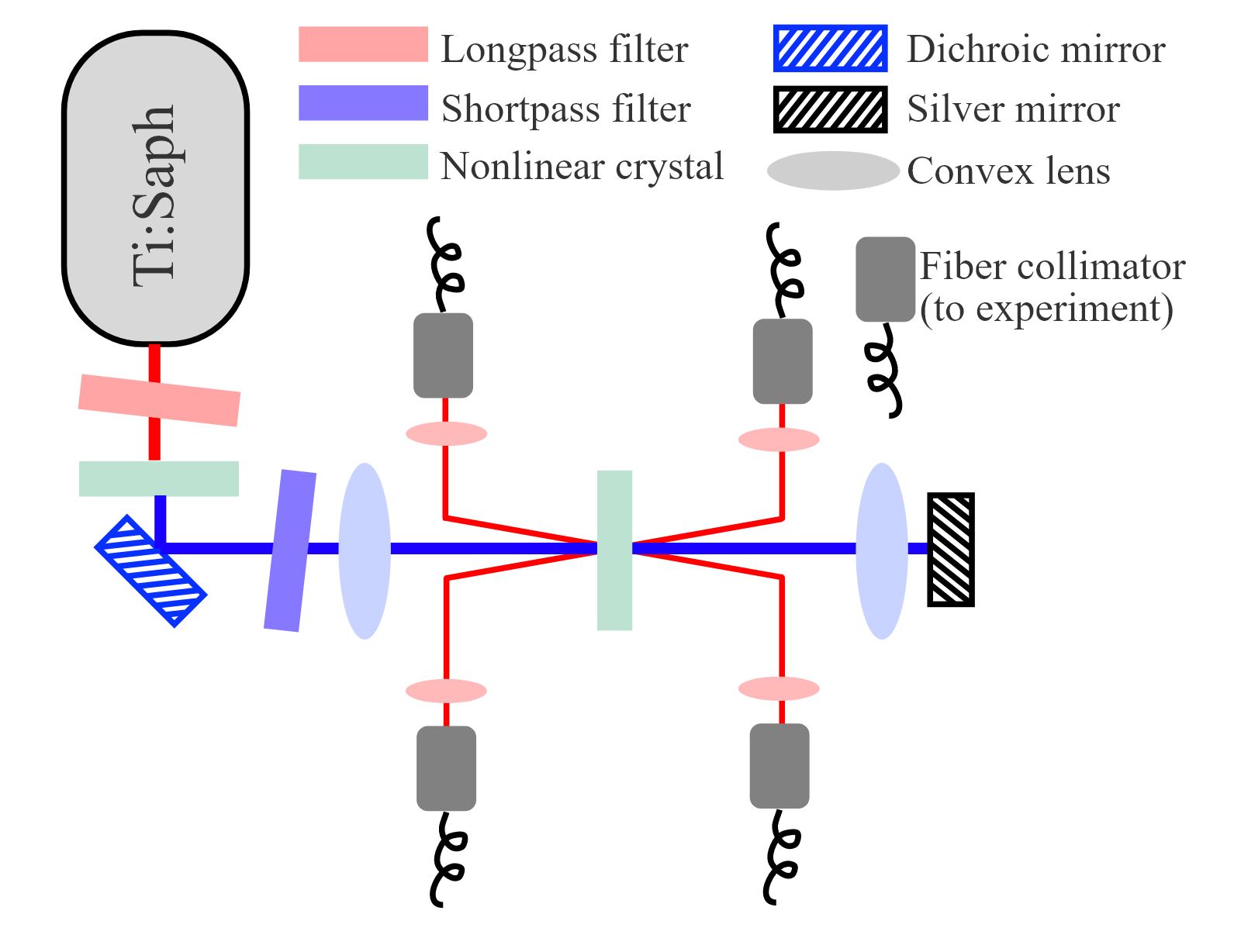}\caption{Source schematic\label{fig:Source-schematic}}
    \end{figure}
}

\newcommand{\figTellipsoid}{
    \begin{figure}[H]
        \begin{centering}
        \includegraphics[width=8.3cm]{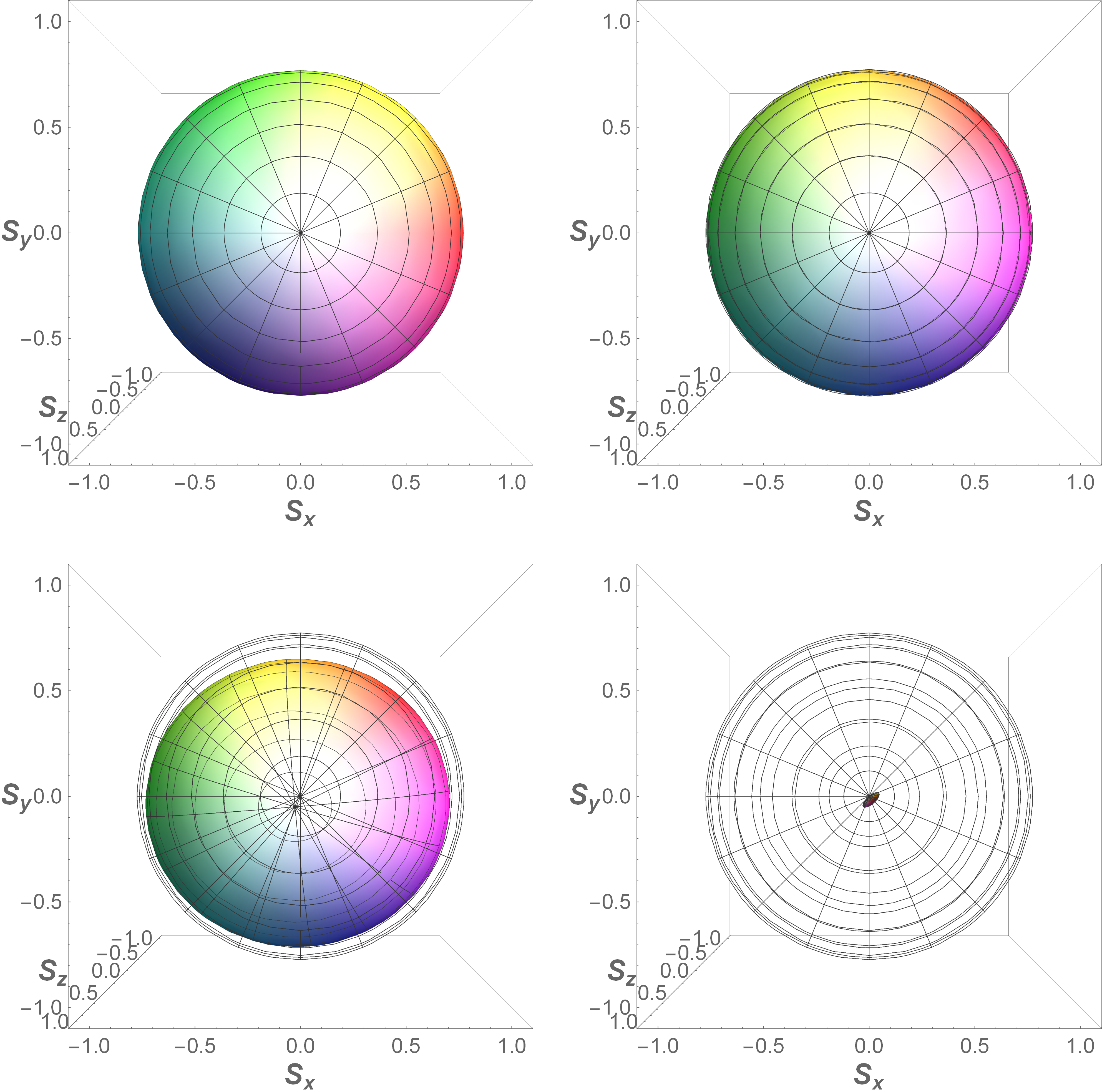}
        \par\end{centering}
        \caption{Tomographic reconstruction of $T$ unitary. From top-left to bottom-right:
        Initial Bloch sphere of pure states; simulation of that Bloch sphere
        under ideal $T$; experimental reconstruction with correct decryption;
        and experimental reconstruction with bad decryption.\label{fig:Tdat}}
    \end{figure}
}

\newcommand{\figTHellipsoid}{
    \begin{figure}[H]
        \begin{centering}
        \includegraphics[width=8.3cm]{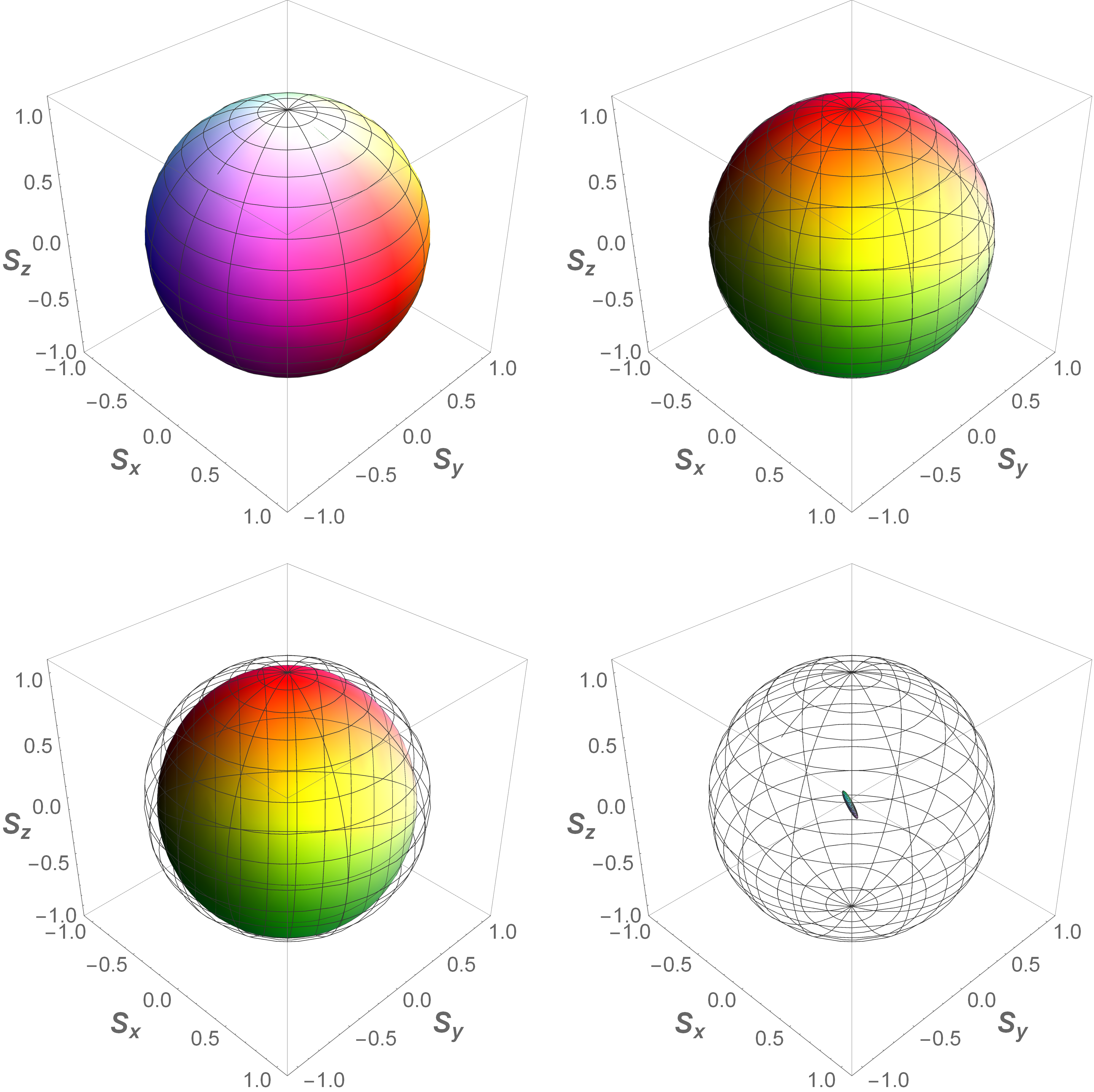}
        \par\end{centering}
        \caption{Tomographic reconstruction of $TH$ unitary. From top-left to bottom-right:
        Initial Bloch sphere of pure states; simulation of that Bloch sphere
        under ideal $TH$; experimental reconstruction with correct decryption;
        and experimental reconstruction with bad decryption.\label{fig:THdat}}
    \end{figure}
}

\newcommand{\figTHPellipsoid}{
    \begin{figure}[H]
        \begin{centering}
        \includegraphics[width=8.3cm]{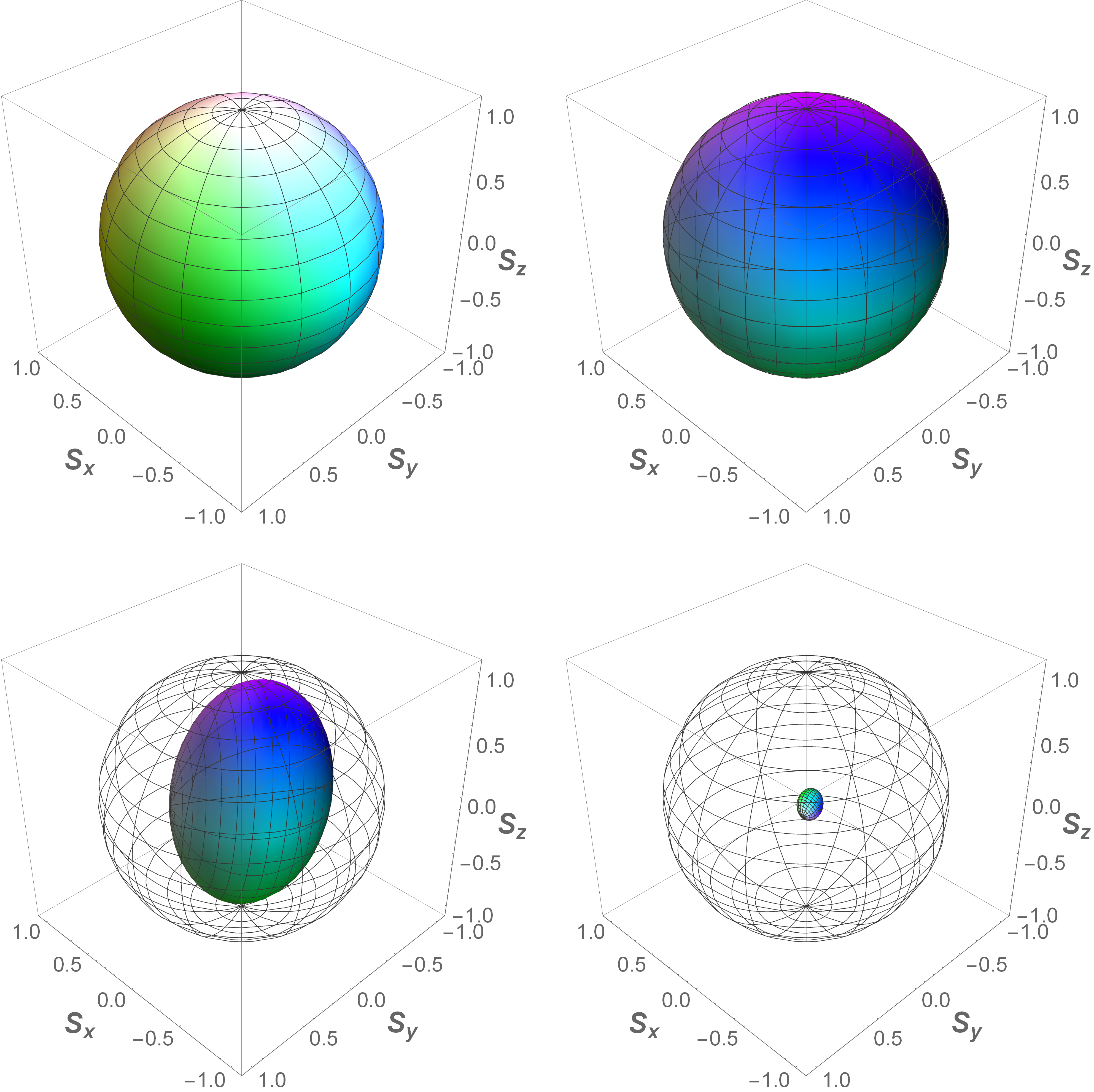}
        \par\end{centering}
        \caption{Tomographic reconstruction of $THP$ unitary. From top-left to bottom-right:
        Initial Bloch sphere of pure states; simulation of that Bloch sphere
        under ideal $THP$; experimental reconstruction with correct decryption;
        and experimental reconstruction with bad decryption.\label{fig:THPdat}}
    \end{figure}
}

\newcommand{\figTHPellipsoidBGsub}{
    \begin{figure}[H]
        \begin{centering}
        \includegraphics[width=8.3cm]{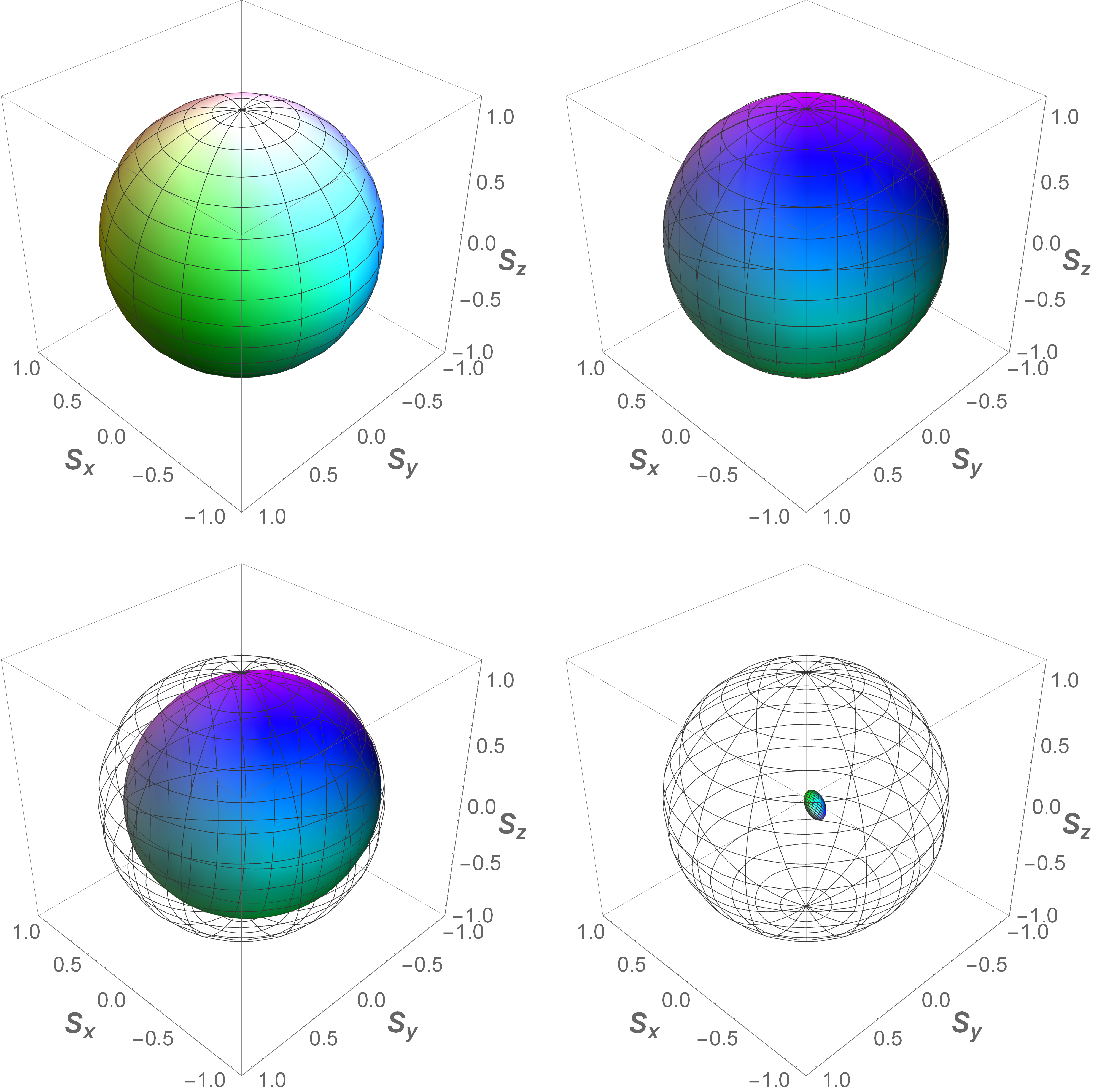}
        \par\end{centering}
        \caption{Background compensated version of Figure \ref{fig:THPdat}.\label{fig:THPdat-BG}}
    \end{figure}
}

\newcommand{\figTbar}{
    \begin{figure}[H]
        \begin{centering}
        \includegraphics[width=8.5cm]{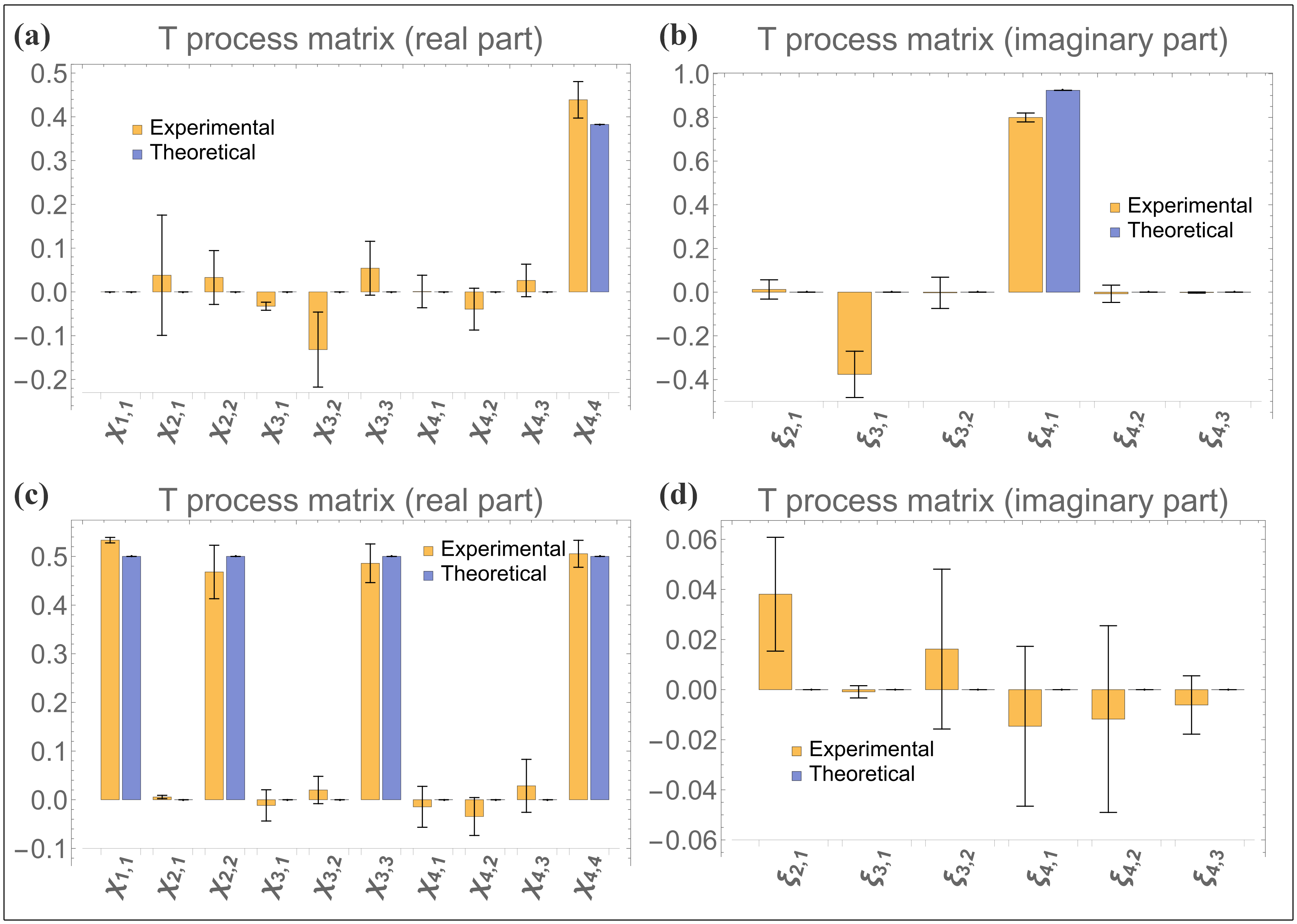}
        \par\end{centering}
        \caption{Tomographic reconstruction of $T$ unitary. From top-left to bottom-right:
        (a) \& (b) real and imaginary parts of process matrix, given \emph{correct}
        decryption; (c) \& (d) real and imaginary parts of process matrix,
        given \emph{wrong} decryption.\label{fig:Tbar}}
    \end{figure}
}

\newcommand{\figTHbar}{
    \begin{figure}[H]
        \begin{centering}
        \includegraphics[width=8.5cm]{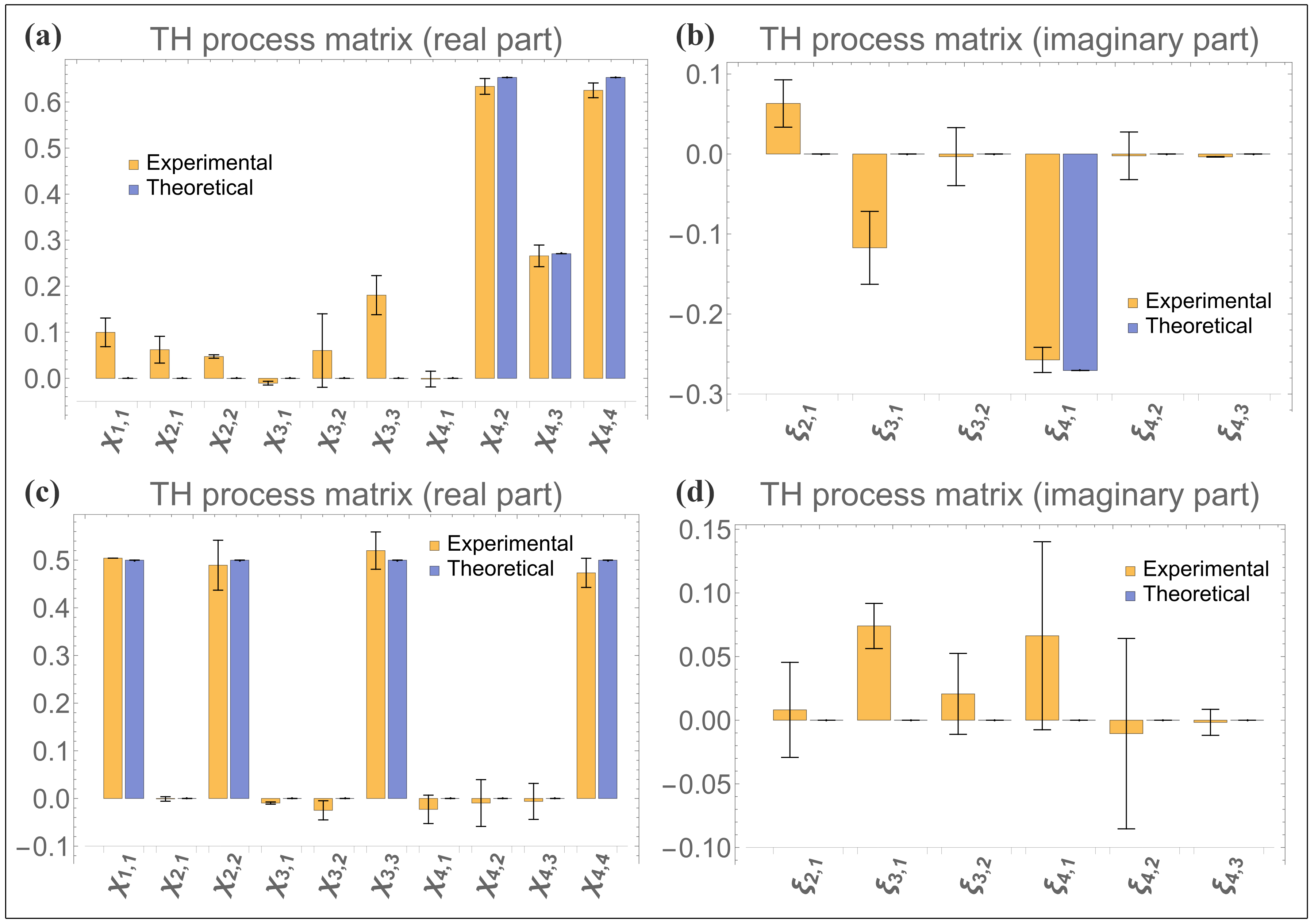}
        \par\end{centering}
        \caption{Tomographic reconstruction of $TH$ unitary. From top-left to bottom-right:
        (a) \& (b) real and imaginary parts of process matrix, given \emph{correct}
        decryption; (c) \& (d) real and imaginary parts of process matrix,
        given \emph{wrong} decryption.}
    \end{figure}
}

\newcommand{\figTHPbar}{
    \begin{figure}[H]
        \begin{centering}
        \includegraphics[width=8.5cm]{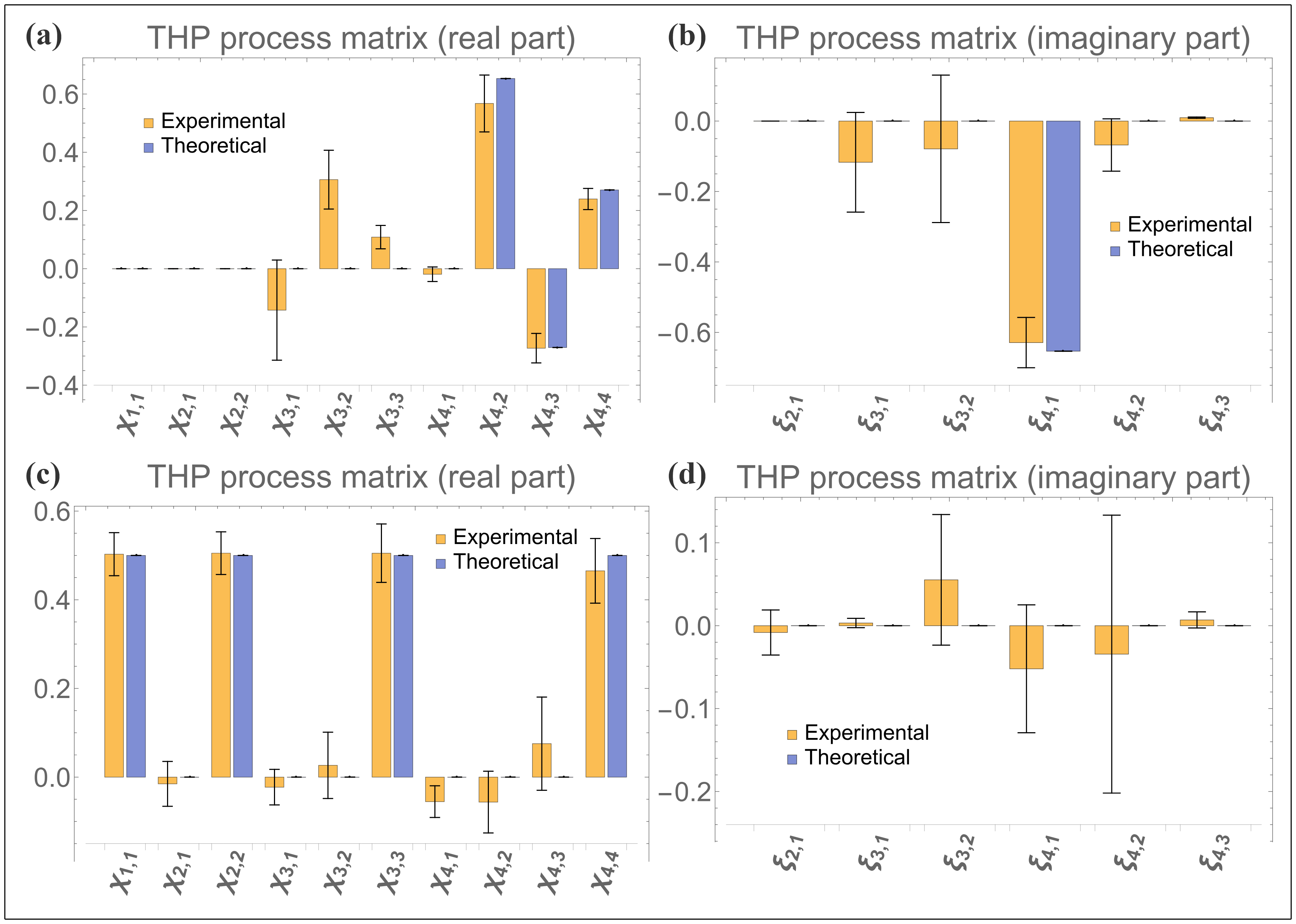}
        \par\end{centering}
        \caption{Tomographic reconstruction of $THP$ unitary. From top-left to bottom-right:
        (a) \& (b) real and imaginary parts of process matrix, given \emph{correct}
        decryption; (c) \& (d) real and imaginary parts of process matrix,
        given \emph{wrong} decryption.\label{fig:THPbar}}
    \end{figure}
}

\newcommand{\figExperimentTPSC}{
    \begin{figure}
        \begin{centering}
        \includegraphics[width=8cm]{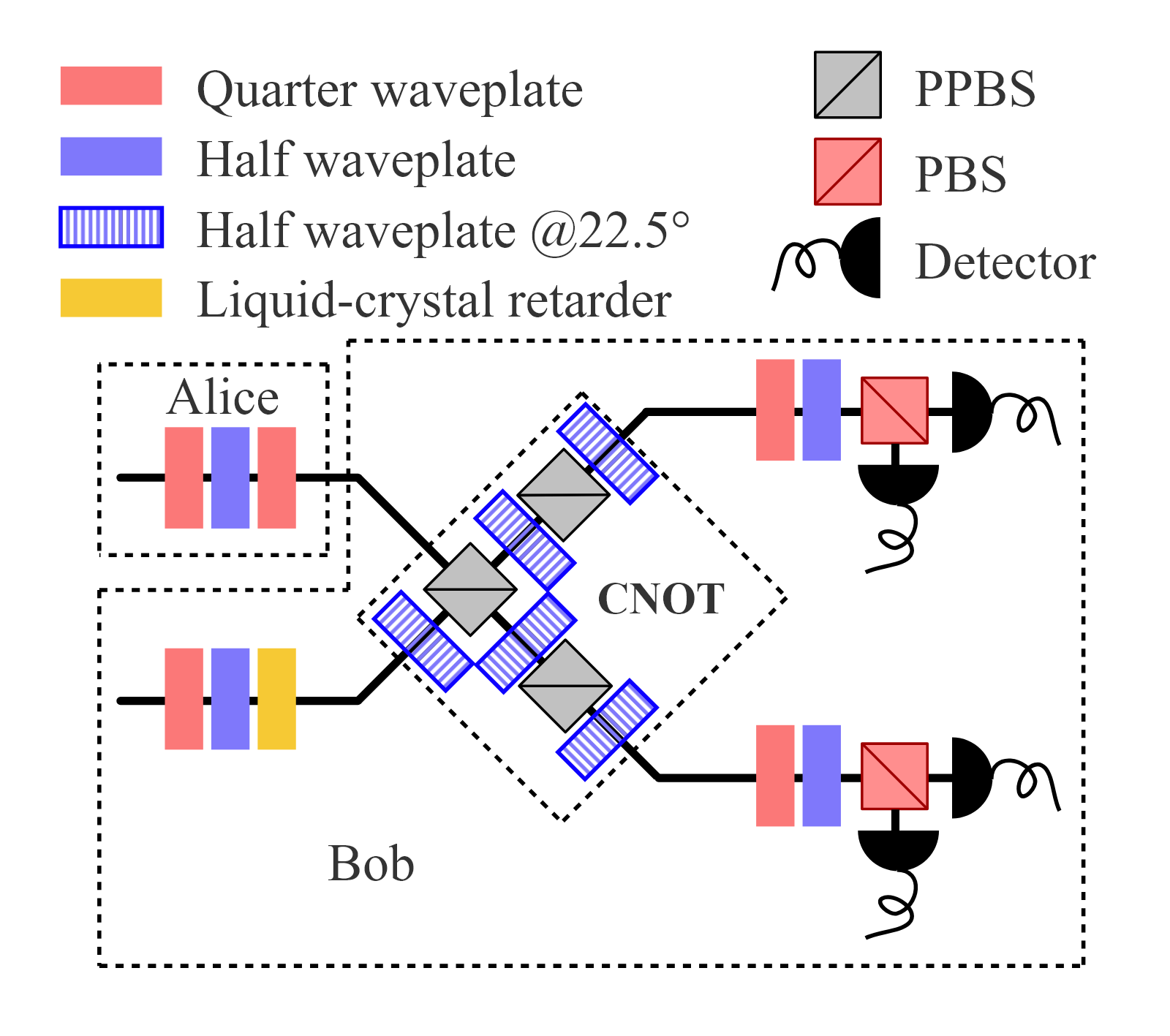}
        \par\end{centering}
        \caption{Experimental setup for our secure comparator two-party protocol.\label{fig:ExperimentalComparator}}
    \end{figure}
}

\newcommand{\figTPSCdataGood}{
    \begin{figure}[H]
        \begin{centering}
        \includegraphics[width=8cm]{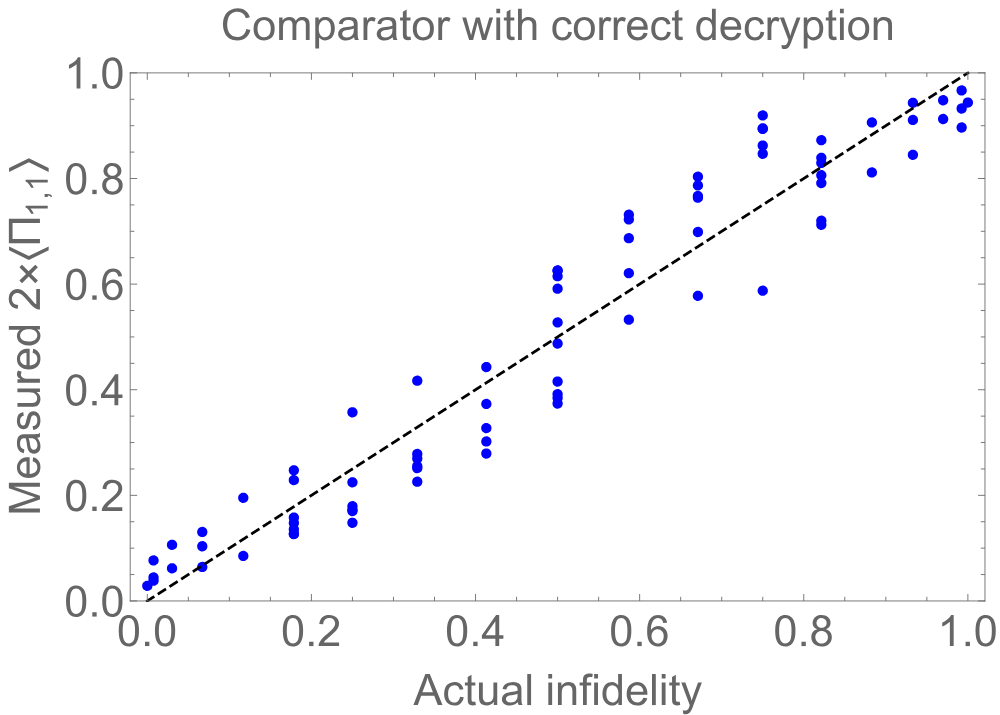}
        \par\end{centering}
        \caption{Plot of measured vs actual infidelity between two states, given correct
        decryption. Dashed line indicate expected theoretical values.\label{fig:2PSC-correct}}
    \end{figure}
}

\newcommand{\figTPSCdataBad}{
    \begin{figure}[H]
        \begin{centering}
        \includegraphics[width=8cm]{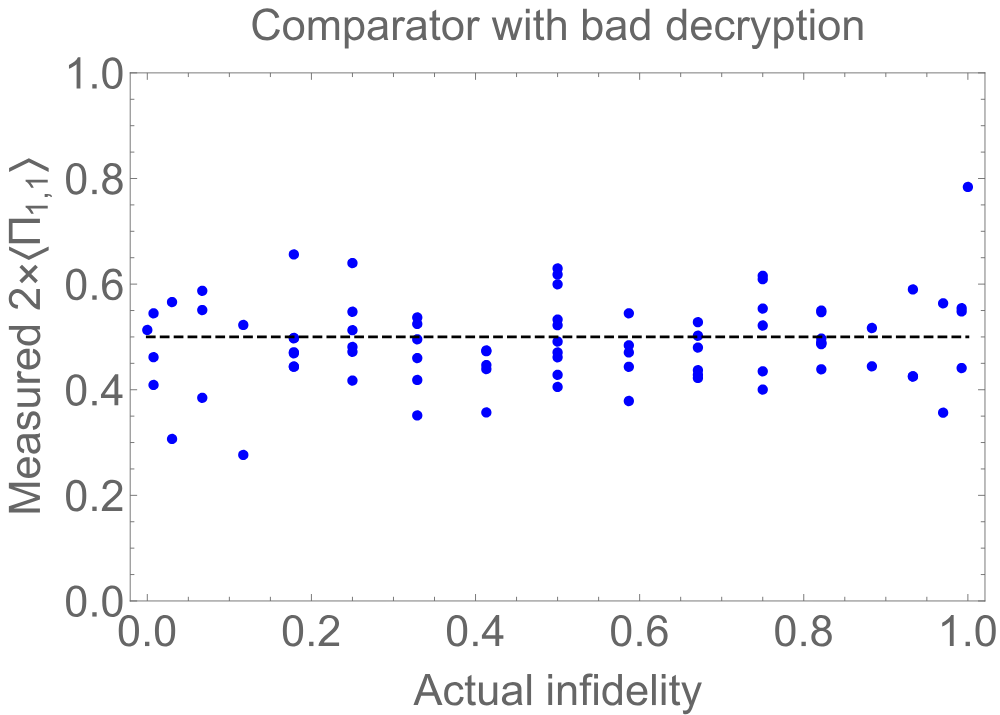}
        \par\end{centering}
        \caption{Plot of measured vs actual infidelity between two states, given wrong
        decryption. Dashed line indicate expected theoretical values -- decrypting with an erroneous key should yield $0.5$ (i.e. infidelity relative to the maximally mixed state). Good agreement with this dashed line indicates that result of the computation is well hidden from parties without a valid decryption key.\label{fig:2PSC-wrong}}
    \end{figure}
}

\begin{document}
\title{Experimental Demonstration of Quantum Fully Homomorphic Encryption\\
with Application in a Two-Party Secure Protocol}

\author{W.K. Tham}
\email{Corresponding author.\\wtham@physics.utoronto.ca}
\author{Hugo Ferretti}
\author{Kent Bonsma-Fisher}
\affiliation{Department of Physics, University of Toronto, 60 St George St, Toronto, Ontario, M5S 1A7, Canada}
\affiliation{Center for Quantum Information and Quantum Control, University of Toronto, 60 St George St, Toronto, Ontario, M5S 1A7, Canada}
\author{Aharon Brodutch}
\affiliation{Department of Physics, University of Toronto, 60 St George St, Toronto, Ontario, M5S 1A7, Canada}
\affiliation{Center for Quantum Information and Quantum Control, University of Toronto, 60 St George St, Toronto, Ontario, M5S 1A7, Canada}
\affiliation{The Edward S. Rogers Department of Electrical and Computer Engineering, University of Toronto, 10 King’s College Road, Toronto, Ontario M5S 3G4, Canada}
\author{Barry C.\ Sanders}
\affiliation{Institute for Quantum Science and Technology, University of Calgary, Alberta T2N 1N4, Canada}
\affiliation{Institute for Quantum Information and Matter, California Institute of Technology, Pasadena, California 91125, USA}
\affiliation{Program in Quantum Information Science, Canadian Institute for Advanced Research, Toronto, Ontario M5G 1M1, Canada}
\author{Aephraim M.\ Steinberg}
\affiliation{Department of Physics, University of Toronto, 60 St George St, Toronto, Ontario, M5S 1A7, Canada}
\affiliation{Center for Quantum Information and Quantum Control, University of Toronto, 60 St George St, Toronto, Ontario, M5S 1A7, Canada}
\author{Stacey Jeffery}
\affiliation{QuSoft and CWI, Amsterdam, the Netherlands}

\begin{abstract}
A fully homomorphic encryption system hides data from unauthorized parties, while still allowing them to perform computations on the encrypted data. Aside from the straightforward benefit of allowing users to delegate computations to a more powerful server without revealing their inputs, a fully homomorphic cryptosystem can be used as a building block in the construction of a number of cryptographic functionalities.
Designing such a scheme remained an open problem until 2009, decades after the idea was first conceived, and the past few years have seen the generalization of this functionality to the world of quantum machines.
Quantum schemes prior to the one implemented here were able to replicate \emph{some} features in particular use-cases often associated with homomorphic encryption but lacked other crucial properties, for example, relying on continual interaction to perform a computation or leaking information about the encrypted data.
We present the first experimental realisation of a quantum fully homomorphic encryption scheme.
We further present a toy two-party secure computation task enabled by our scheme.
Finally, as part of our implementation, we also demonstrate a post-selective two-qubit linear optical controlled-phase gate with a much higher post-selection success probability (1/2) when compared to alternate implementations, e.g.\ with post-selective controlled-$Z$ or controlled-$X$ gates (1/9).
\end{abstract}
\maketitle

\section{\label{sec:Intro}Introduction}
In 1978, Rivest, Adleman, and Dertouzos first imagined constructing a cryptosystem with the property that a party \emph{without} a valid secret key required for decryption can nevertheless correctly evaluate a function $f$ directly on a ciphertext $x$, without learning anything about either $f(x)$ or $x$~\cite{rivest1978data}.
In addition to the obvious benefit of being able to delegate computation to a party that is otherwise not trusted with private data, cryptographers have observed that elegant cryptographic solutions to particularly interesting tasks can be constructed on top of a fully homomorphic encryption scheme--secure multiparty computation, non-interactive zero-knowledge proofs, one-time programs, to name a few~\cite{Goldwasser2PSCpoker,Goldreich2PSC,GoldreichObliviousRam,GoldwasserOTP,KilianZKP}.
Despite the apparent utility of such an encryption scheme, the question of whether it was possible to efficiently construct one remained open until 2009 when the first fully homomorphic encryption (FHE) scheme was constructed for classical machines~\cite{homenc}.

In quantum computing, a range of works have addressed the closely related problem of secure delegated computing~\cite{ChildsSecureAssistedQC,BroadbentOTP,UBQCtheory,ABE10,Bro15,DFPR13,UBQCexp,KentComputingonEncryptedData} wherein Alice (whose quantum computer may be of limited power) can implement a quantum computation with the help of Bob (who possesses a more powerful quantum computer), without revealing her computation (or equivalently, her input). In these secure delegated quantum computing schemes, interaction between Alice and Bob can occur repeatedly as they collaborate to perform the desired computation. By contrast, in a fully homomorphic setting Bob is allowed to apply any quantum computation of his choice without further assistance from Alice. A secure delegated computing scheme therefore is not amenable to the wider gamut of cryptographic uses that a true FHE scheme should be, even if they are similar in spirit insofar as both allow delegation of computation on private data.

The question of quantum fully homomorphic encryption was studied more recently. The first proposals treated the problem in an information-theoretic security setting, where QFHE was subsequently shown to be impossible unless severe compromises to universality or security were made~\cite{YPDF14}. Schemes that were proposed under this model~\cite{RFG12,TKO14}, and subsequently demonstrated experimentally~\cite{WaltherHomEnc}, leaked parts of the input and therefore do not satisfy mainstream notions of cryptographic security. Under standard cryptographic security, analogous to classical FHE where security necessarily requires computational assumptions, two QFHE schemes were theoretically proposed for tasks with a limited number of $T$ gates~\cite{StaceyQFHE}. This work left open the question of QFHE for arbitrary quantum circuits, but could be seen as the first QFHE scheme for circuits with a small number of $T$ gates. A theoretical extension later removed this limitation to enable QFHE on circuits with an arbitrary number of $T$ gates~\cite{SpeelmanCompactQFHE}.

In this work, we implement and experimentally demonstrate for the first time a QFHE scheme proposed in~\cite{StaceyQFHE}. In addition to the core scheme, we also experimentally demonstrate a small two-party task enabled by QFHE that cannot otherwise be performed. In this task Alice and Bob are to compare and compute the inner product between their qubit states, without being able to learn any additional information (e.g. tomographically complete statistics) about the other party's state. Various two- and three-qubit circuits in these demonstrations are implemented optically in a four-photon setup, with each qubit encoded in photon polarisation and one photon serving as herald. Our experiment includes a novel realisation of a phase-add gate, a two-photon operation with post-selection success probability of 1/4; far higher than 1/9 for an equivalent implementation with a more typical post-selective two-qubit gate in photon polarisation.

The rest of this paper is structured as follows. In section \ref{sec:Theory} we lay out various attributes required in a QFHE scheme followed by a detailed description of the protocol that we implemented. Section \ref{sec:Experiment} describes our experimental apparatus and then proceeds to discuss data showing the core QFHE scheme at work. Finally, section \ref{sec:2PSC} details the two-party computation task and discusses experimental data pertaining to it.

\section{\label{sec:Theory}Theory background}
\subsection{\label{sec:DefQFHE}What makes an encryption scheme homomorphic?}
A public key encryption scheme describes a procedure ${\cal E}_{pk}$ for encrypting a plaintext $\phi$ using a public key, $pk$, to get a ciphertext ${\cal E}_{pk}(\phi)$; as well as a procedure ${\cal D}_{sk}$ for decrypting a ciphertext $\psi={\cal E}_{pk}(\phi)$ to recover the plaintext $\phi={\cal D}_{sk}(\psi)$ using the secret key $sk$.

A \emph{homomorphic} encryption scheme derives its name from the fact that, \emph{operationally}, it behaves like a homomorphism between plaintext and ciphertext (call these $\phi$ and $\psi$ respectively). Loosely speaking, each valid operation (e.g.\ modular addition/multiplication) on $\phi$ maps to a well-defined operation on the corresponding $\psi$, called an `evaluation map'. If we write an operation between plaintext as $\diamond$, and the corresponding operation (or evaluation map) between ciphertext as $\circ$, we want: ${\cal E}_{pk}(\phi_1\diamond \phi_2)\equiv {\cal E}_{pk}(\phi_1)\circ {\cal E}_{pk}(\phi_2)=\psi_1\circ\psi_2$ for any $\phi_1$ and $\phi_2$.

While the description above serves as a useful template, no fully homomorphic cryptosystem in practice actually works by leveraging homomorphisms in $\mathcal{E}_{pk}$ and $\mathcal{D}_{sk}$ in the strict sense. Instead, the evaluation map for an operation $\diamond$ takes, as input, ciphertext $\psi_1$ and $\psi_2$ and outputs a ciphertext that would decrypt to $\phi_1\diamond \phi_2$ -- not necessarily the same ciphertext you would get from encrypting $\phi_1\diamond\phi_2$ directly.
On a quantum machine, $\ket{\phi}$ and $\ket{\psi}$ are states in a Hilbert space acted upon by quantum channels, for example, a unitary gate, or sequence of gates, $U$. By analogy, for quantum homomorphic encryption, the evaluation map for $U$ is a quantum channel $U'$ with the property: ${\cal D}_{sk} U' {\cal E}_{pk}\ket{\phi} = U\ket{\phi}$. 

To be considered fully homomorphic, we require an encryption scheme to satisfy certain properties:

\begin{itemize}
    \item There must be an efficiently computable evaluation map for \emph{any} valid operation on the plaintext. In a gate-based quantum computing model this means \emph{every} gate in a universal gateset (e.g. Clifford + T). A scheme that is only partially homomorphic (RSA cryptosystem~\cite{RivestRSA}, famously homomorphic only in multiplication modulo $N$) doesn't lend itself to many use-cases often associated with FHE.
    \item A party (say Bob) in possession of $\psi$ ought be able to perform the evaluation map for an arbitrary sequence of gates $U$ of his choice without further assistance from other parties (say Alice). In practice, we augment this by allowing Alice to supply a combination of specially prepared qubits and classical data collectively called the `evaluation key'. Crucially, these should be generated \emph{at the same time} as Alice prepares $\psi$, and must not themselves depend on $U$.
    \item The scheme should be compact, meaning that the difficulty of decrypting the output of the evaluation map for $U$ should not depend on the difficulty of computing $U$. At the very least, decryption should not be as difficult as computing $U$, otherwise homomorphic evaluation becomes trivial: simply append a \emph{description} of $U$ to a ciphertext, and leave Alice to apply it as part of the decryption procedure.
\end{itemize}

\subsection{\label{sec:PartialScheme}An almost-homomorphic scheme}
\figCliffordKeyTransforms
To construct a QFHE scheme, we begin with the secure delegated computation scheme in~\cite{KentComputingonEncryptedData}, which we briefly describe here.
Alice generates random classical bits $a,b\in\left\{ 0,1\right\} $ and encrypts her qubit by applying corresponding Pauli operators: $\ket{\psi}=Z^{a}X^{b}\ket{\phi}$.
If $a$ and $b$ are drawn from a uniform distribution and used as one-time-pads, $\ket{\psi}$ looks like the maximally mixed state to everyone but Alice.
Bob then proceeds to compute on this ciphertext with some unitary $U'$ to produce $U'\left|\psi\right>$.

If $U'$ consists only of a sequence of Clifford gates then Alice can decrypt and obtain the correct (computed) plaintext again by applying the correct Pauli operators: $Z^{a'}X^{b'}U'\left|\psi\right>=U'\left|\phi\right>$.
As long as Bob's gate sequence $U'$ is known to Alice, she can deduce the correct decryption keys ($a'$ and $b'$) from her encryption keys ($a$ and $b$) (see Table~\ref{tab:CliffordKeyTransform}).
    
A non-Clifford element, like the $T$ gate ($T=\left|0\right>\left<0\right|+e^{i\pi/4}\left|1\right>\left<1\right|$), must be added to the Clifford set for quantum computation to be universal. However, as soon as Bob performs a $T$ gate, decryption requires more than just Pauli operators.
As shown in Fig.~\ref{fig:TgatePcorrection} a phase gate that isn't otherwise part of $U'$ \emph{may} need to be applied, depending on encryption key $b$ and all Clifford operators that precede the $T$ gate.

Performing the phase correction without requiring Alice to divulge encryption key $b$ to Bob can be done with an ancilla qubit containing the correct phase. Each time Bob performs a $T$ gate, he communicates to Alice all operations performed prior to that $T$ gate. Alice prepares an ancilla with an appropriate phase. Bob uses this ancilla in the simple teleportation circuit in Fig.~\ref{fig:figSemiSimpleT}, which allows that phase (modulo $\pi$) to be `kicked back' onto $\left|\psi\right>$ thereby performing the desired post-$T$-gate phase correction.
\figTsimple

Shortcomings in this almost-homomorphic protocol are clear. Primarily, it is an interactive one: every time Bob performs a $T$ gate, he must communicate with Alice. Both quantum (ancilla qubit) and classical (Bob's description of $U'$) data must be exchanged during each interaction. More critically, preparation of ancilla and decryption of ciphertext can only be done if Alice knows Bob's computation, $U'$. Neither of these attributes is desired in a QFHE scheme.

\subsection{\label{sec:OurScheme}A QFHE scheme}
The scheme implemented in this paper, which was first proposed and theoretically analysed in~\cite{StaceyQFHE}, makes the following crucial modifications to the protocol just described above to make it homomorphic:
\begin{itemize}
    \item First, we `front-load' preparation of all ancilla qubits. During encryption Alice prepares two ancilla qubits: $\left|\xi_{a}\right>=Z^{q}P^{a}\left|+\right>$ and $\left|\xi_{b}\right>=Z^{r}P^{b}\left|+\right>$ where $q,r\in\left\{ 0,1\right\}$ is another set of random bits. Observe from Table~\ref{tab:CliffordKeyTransform} that single-qubit Clifford gates can transform each decryption key into just three possible values: they become either $a$, $b$, or $a+b$. If the Clifford sequence preceding a $T$ gate implies $b'=a$ or $b'=b$, Bob uses $\left|\xi_{a}\right>$ or $\left|\xi_{b}\right>$ respectively in a teleportation circuit (Fig.~\ref{fig:figSemiSimpleT}) to apply the require phase correction. But if $b'=a+b$, Bob first applies that teleportation between the ancillas to obtain: $Z^{s}P^{a+b}\left|+\right>$, which has the correct phase. This resulting qubit is then used as before to correct the phase on $\left|\psi\right>$ (see Table~\ref{tab:T-correction}). If additional $T$ gates are anticipated in the gate sequence, Alice simply prepares more ancillas correspondingly.
    \item Second, we leverage a \emph{classical} FHE scheme to make our quantum protocol truly homomorphic.
    As she prepares ciphertext $\ket{\psi}$, Alice also homomorphically encrypts her encryption keys: $a\to\text{Enc}\left(a\right)$ and $b\to\text{Enc}\left(b\right)$, and sends them to Bob. During evaluation, as he computes $U'\ket{\psi}$, Bob also transforms these encrypted classical keys correspondingly: $\ensuremath{\text{Enc}\left(a,b\right)\to\text{Enc}\left(a',b'\right)}$. We stress that Bob can do this only because the (classical) encryption on $a$ and $b$ is fully homomorphic.
\end{itemize}

These modifications together address shortcomings in the almost-homomorphic scheme that we've described. The first modification makes the protocol non-interactive. All resources that Bob needs to evaluate his circuit correctly, including ancilla qubits, are prepared and sent at the beginning of the protocol. Combining the protocol with classical FHE negates the need for Alice to be cognizant of Bob's circuit -- the homomorphically encrypted secret key, suitably modified by Bob during computation, allows her (and only her) to correctly decrypt $\ket{\psi}$. In section \ref{sec:2PSC} we discuss a use-case in which these features are important.

No protocol is without caveats however. Briefly, the reader should be aware that the QFHE protocol we demonstrate here: a) assumes the existence of a classical FHE that is secure against a quantum adversary and b) requires an evaluation key (i.e. ancilla qubits) whose size grows doubly exponentially with the $T$-depth (i.e. number of layers of $T$ gates \emph{sequentially} applied). A clever way to modify evaluation of $T$ gates in order to solve (b) was proposed in \cite{SpeelmanCompactQFHE}. However, if Bob's circuit is known to have a fixed $T$-depth, as in our implementation, then this is not an issue. With regards to (a), known classical FHE schemes are believed to be resistant to quantum attacks~\cite{bernsteinPostQuantum}.

\figCanonicalCases

\section{\label{sec:Experiment}Experimental realisation}
\figExperimentSchematic
We implemented the core QFHE protocol described above in section \ref{sec:OurScheme} in an optical setup. Our implementation accommodates a total of three (1 data + 2 ancilla) logical qubits and is capable of performing arbitrary single-qubit rotations, but is limited to a single two-qubit gate per pair of qubits and to circuits of $T$-depth one (i.e. no cascaded $T$ gates). These latter restrictions are simply due to the limited scale of the specific setup we constructed in our laboratory, and do not imply any fundamental limitation of the protocol in general.

A key distinguishing feature of our protocol is its handling of encryption/decryption by Alice as well as $T$ gates performed by Bob. The reader will recall that Clifford gates preceding a $T$ gate modify the secret key in just one of three ways, each of which necessitates a different usage of the ancilla qubits. We sought to demonstrate our protocol is both correct and secure when handling each of these three canonical cases involving a $T$ gate. To this end, we realised the three corresponding circuits shown in Table~\ref{tab:T-correction} in our optical implementation.

Qubits in our implementation are encoded in the polarisation degree-of-freedom of photons. Single-qubit gates are realised using standard birefringent polarisation optics.
Two-qubit gates, specifically controlled-$X$ and controlled-$Z$ gates, are implemented post-selectively by leveraging bosonic bunching or the Hong-Ou-Mandel (HOM) effect~\cite{KokLinearOpticsQCreview,HongOuMandel,WeinfurterCphase,AGWhiteCNOT}.
Fig.~\ref{fig:Experimental-setup} shows a schematic of our optical apparatus.\figSource

Photons in our experiment are produced in a type-1 spontaneous parametric down conversion (SPDC) source. The source is a $2$mm thick BBO crystal pumped with $404$nm blue light produced via second harmonic generation (SHG).
The SHG setup is in turn pumped with a Ti:Sapphire laser (Coherent Chameleon Ultra) configured to pulse with a repetition rate of $80$MHz, center wavelength of $808$nm and transform-limited pulse duration of $150$fs. Fig.~\ref{fig:Source-schematic} shows a schematic of our source. During use, photons from this source are spectrally narrowband-filtered ($3$nm FWHM, Edmund Optics). When power for the blue ($404$nm) pump is set to average at $10$mW, the source produces photons at a rate of approximately $5000$~pairs/second with a heralding efficiency of about $16$\% (before losses accrued in optical gates), and 4-fold coincidences of $
0.5$~quads/second. Our detectors are a combination of Perkin-Elmer (now Excelitas) SPCM-AQRH-W2 and SPCM-AQ4C modules coupled to home-built coincidence logic configured to operate with a coincidence window of $4$ns.

The classical FHE that we have selected for encrypting all classical keys is the ``Brakerski-Gentry-Vaikuntanathan'' scheme~\cite{BGValgo} as implemented in the `HElib' library~\cite{HElib}, lightly modified for easy integration with the experiment. The library was called with a plaintext base $p=2$, security parameter $k=80$ and number of plaintext slots $l=150$.

\subsection{\label{sec:AuxGateArch}A novel optical gate architecture for adding phases}
    General two-qubit gates in our experiment are implemented via second order interference of indistinguishable photons at a beamsplitter. If the beamsplitter is carefully designed to fully transmit and not reflect horizontal polarised photons ($T_{H}=1$, $R_{H}=0$), while partially reflecting vertical polarised photons ($T_{V}=1/3$, $R_{V}=2/3$), one can easily show that the effective operation upon post-selection in the coincident basis is a controlled-$Z$ gate in polarisation~\cite{WeinfurterCphase}. In Fig.~\ref{fig:Experimental-setup} we refer to such an optical element as a partially polarising beamsplitter (PPBS). We note that the post-selection success probability here is $1/9$.

Observe, however, that it is not always necessary to implement a general controlled-$X$. The controlled-$X$ between ancillas in case~3 of Table~\ref{tab:T-correction}, for example, merely serves as a means of accomplishing phase addition (modulo $\pi$) between $\left|\xi_{b}\right>$ and $\ket{\xi_a}$. That is, we want a channel such that: $\left(\left|0\right>+e^{i\phi_{a}}\left|1\right>\right)_{a}\otimes\left(\left|0\right>+e^{i\phi_{b}}\left|1\right>\right)_{b}\to\left(\left|0\right>+e^{i\left(\phi_{a}+\phi_{b}\right)}\left|1\right>\right)_{a}$.

However, since both ancillas are confined to states on the equator of the Bloch sphere, it turns out we can be more efficient by replacing the PPBS with a fully polarising beamsplitter (PBS, i.e., $T_{V}=0$, $R_{V}=1$). Consider the mode transformation of the PBS followed by a Hadamard gate
(i.e., halfwave plate at $22.5$~deg) on one output arm, acting on ancilla states $\ket0+e^{i\alpha}\ket1$ and $\ket0+e^{i\beta}\ket1$ (in our convention, a photon in $H$ or $V$ polarisation encodes $\ket0$ or $\ket1$ respectively):\eqAuxHOM
where $\hat{a}^\dagger$ and $\hat{b}^\dagger$ are bosonic creation operators in the two input/output modes of the PBS. Subscripts on these operators label polarisation.
The Hadamard gate following the PBS acts on mode $b$. In the last line, we've omitted terms that do not contribute to coincidence events between modes $a$ and $b$. Now post-selecting on finding a photon in either $H$ or $V$ (i.e. $\ket0$ or $\ket1$) in mode $b$, we find the photon in mode $a$ is left in state $\ket{\zeta}$:\eqAuxPS
    
Note that $\ket{\zeta}$,
when post-selection on $\hat{b}_\text{V}^\dagger$ succeeds,
is exactly the qubit state that we expect on the middle rail in case~3 of Table~\ref{tab:T-correction}, when classical bit $k_{1}=0$.
If post-selection on $\hat{b}_\text{H}^\dagger$ succeeds on the other hand, $\ket{\zeta}$ does not quite correspond
to $k_{1}=1$. Nevertheless, $\ket{\zeta}$ carries the correct phase modulo $\pi$ so all that is required is that we update the key transformation rule indicated in Table~\ref{tab:T-correction}. That rule should now read:\eqAuxNewKey
where we now define $k_{1}=0$ when post-selection on $\hat{b}_\text{V}^\dagger$ succeeds and $k_{1}=1$ when $\hat{b}_\text{H}^\dagger$ succeeds.
    
We also note that the success probability for post-selection of each polarisation on the $b$ mode photon is $1/4$, so that the total post-selection success probability is $1/2$. This is far more efficient than $1/9$, which we expect from a general controlled-$X$.

\subsection{\label{sec:DatnRes}Data and results}
With the optical apparatus shown in Fig.~\ref{fig:Experimental-setup} we implemented the circuits in Table~\ref{tab:T-correction}. These were designed to highlight our protocol's novel aspects under three canonical secret key transformations (discussed in section~\ref{sec:OurScheme}) due to Cliffords preceding a $T$ gate; $T\ket{\psi}$, $TH\ket{\psi}$, $THP\ket{\psi}$ being the simplest of these.

Alice prepares and encrypts her data qubit in the photon on the top-left rail. She also prepares two appropriate ancillas encoded in photons on the two bottom-most rails. When implementing cases 1 and 2 (i.e. when Bob evaluates $T\ket{\psi}$ or $TH\ket{\psi}$), the phase-add gate is not required so we swap the appropriate ancilla up into the second rail, where it is then allowed to interact with the ciphertext qubit at ``CNOT 1''.

When implementing case~3 (i.e. when Bob evaluates $THP\ket{\psi}$), phases on both ancillas are first summed at the phase-add gate. Meanwhile, the ciphertext qubit is allowed to bypass ``CNOT 1'' (relevant optical components are moved out of that photon's path). The ciphertext qubit that began as the top-left photon now propagates directly to ``CNOT 2'', where it is entangled with the remaining ancilla (the other now serves as a herald for successful operation of the phase-add gate).

    All (classical) bits resulting from measurement that Bob performs while evaluating $T\ket{\psi}$, $TH\ket{\psi}$, or $THP\ket{\psi}$ are sent back to Alice in order that she be able to perform decryption correctly. To verify that the protocol works as advertised, Alice prepares a variety of plaintext states and measures in a variety of bases after decryption so as to be able to infer the process map for Bob's evaluation. If indeed the protocol is correct,
    Alice's tomographic reconstruction of Bob's process should closely match the ideal $T$, $TH$, or $THP$ unitary.
    
    We further repeat the experiment, this time with Alice's secret keys purged and replaced with a set of erroneous keys before she has a chance to decrypt ciphertexts returned by Bob. In this case, we expect the tomographically reconstructed process to be the completely depolarising channel instead. Decryption with erroneous keys is accomplished by asking HElib to generate two sets of keys, $sk_{1}$ and $sk_{2}$, and programming Alice to compute $\text{Dec}_{sk_{2}}\left(\text{Enc}_{sk_{1}}\left(a,b,i,j\right)\right)$, thereby simulating what an attacker with no access to the correct key $sk_{1}$ might observe.

    Figures \ref{fig:Tdat}, \ref{fig:THdat}, and \ref{fig:THPdat} show plots of how a unit (Bloch) sphere transforms under each canonical gate sequence enumerated in Table~\ref{tab:T-correction}. The top-left panel in each figure shows a unit Bloch sphere representing the set of all possible input plaintext state, each represented by a unique color. Top-right panels in turn show what each state maps to (ideally) under the $T$, $TH$, or $THP$ gate sequence. For comparison, the bottom-left panels show tomographic reconstructions of Bob's process when decryption \emph{is} done correctly---these should resemble the top-right panels. Finally, bottom-right panels show tomographic reconstructions of Bob's process when decryption \emph{is} \emph{not} done properly -- these should resemble the fully depolarising channel.
    
    We compare our experimentally reconstructed channels with an ideal desired channel by calculating the average process fidelity~\cite{MolmerProcessFidelity}. When decryption is done correctly, these process fidelities are $96.1\%$, $96.2\%$, and $83\%$ with respect to the ideal $T$, $TH$, and $THP$ unitary respectively. When decryption is \emph{not} done correctly, the corresponding process fidelites are $51.7\%$, $47.4\%$, and $50.2\%$ (the fully depolarising channel has process fidelity of $50\%$ with respect to \emph{any} unitary). We also compare these latter channels directly with the ideal fully depolarising channel, yielding process fidelities of $99.8\%$, $99.7\%$, and $99.4\%$ respectively, suggesting that any attempt to decrypt the ciphertext with an erroneous key will only result in noise.

    In the case of correct decryption, the primary limiting factor for these experimental process fidelities is the quality of our two-qubit gates. In turn, the dominant source of error in these gates is imperfect two-photon interference visibility. Photons from the same SPDC event (i.e., ``intra-pair'') exhibited HOM interference contrast of $97.0\pm0.5\%$ at a 50/50 BS ($\sim77\%$ at a PPBS, where $80$\% is expected) whereas photons from different SPDC events (i.e., ``inter-pair'') had a contrast of $90.0\pm1.5\%$ ($\sim72\%$ at a PPBS). Experimental fidelity for the $THP$ unitary is significantly worse than the other two because of the need to cascade two two-qubit gates, thereby compounding errors. Specifically, the phase-add gate is followed by ``CNOT 2'' as described above and shown in Fig.~\ref{fig:Experimental-setup}, the latter operating by HOM interference between photons from different SPDC events.

    \figTellipsoid \figTHellipsoid \figTHPellipsoid \figTHPellipsoidBGsub

    Recall that our experimental two-qubit gates yield the correct processes when second-order interference between two single-mode bosonic creation operators occurs at a beamsplitter followed by post-selection on exactly one boson in each output mode. Practical limitations in construction of the apparatus (e.g. imperfect alignment of collection modes, defects in collection optics, and variances in spectral profile of narrowband filters) contribute to finite interference contrasts. Furthermore, inter-pair photons may not have the same (coherent) spectral correlations that exist in intra-pair ones, thereby partially invalidating the single-mode assumption.
    
    A less obvious but equally important effect is the contribution of parasitic processes that lead to unwanted coincidence events. For instance, higher-order SPDC events that yield more than one photon per output mode can contribute to successful post-selection (i.e., coincidence) events that do \emph{not }yield the desired output states. Similarly, because our coincidence windows while small are nevertheless finite, two uncorrelated photons or detector noise can nevertheless register erroneously but positively on our coincidence circuit.

    Practical limitations imposed by equipment or procedural imperfections cannot be remedied easily. And while background processes due to higher-order SPDC events can be mitigated by reducing pump power, doing so results in impractically small signals. However, since our apparatus is sufficiently well characterised, we can calculate the expected prevalence of background processes described above and subtract them from our signal in post-processing. Background-subtracted experimental data for the $THP$ unitary is shown in Fig.~\ref{fig:THPdat-BG}.
    With background subtraction, process fidelity for the tomographic reconstruction of the THP unitary with correct decryption increases from $83\%$ to $94\%$.

    In the interest of thoroughness, we have also presented the same data in bar-chart form in Figs.~\ref{fig:Tbar} through \ref{fig:THPbar}. Those figures show the magnitude of elements of the process matrix.
    In the Kraus representation of a qubit map $\rho_{out}=\sum_{j}K_{j}\rho_{in}K_{j}^\dagger$, the matrix $M_{jk}=\chi_{jk}+i\xi_{jk}$ succinctly defines Kraus operators $K_{i}$ in terms of a standard Pauli basis: $K_{j}=\sum_{k}\left(\chi_{jk}+i\xi_{jk}\right)\sigma_{k}$. Here $\chi,\xi\in\mathbb{R}$ and $\sigma_{k}$ is to be interpreted as a Pauli matrix with the following labelling: $\sigma_{0}=I$, $\sigma_{1}=X$, $\sigma_{2}=Y$, and $\sigma_{3}=Z$. Reconstructions were performed using standard maximum likelihood estimation (MLE)~\cite{GambettaTomo}.\figTbar \figTHbar \figTHPbar

\section{\label{sec:2PSC}An application of QFHE: two-party secure computation}
\subsection{\label{sec:2PSCtheory}Protocol Description}
In this section, we describe a protocol we developed in order to demonstrate a use-case for QFHE that is otherwise difficult to accomplish. Imagine Alice and Bob each possesses a qubit state, $\rho_{\alpha}$ and $\rho_{\beta}$ respectively. They are tasked with finding the inner product or fidelity between their states \begin{equation}
\label{eq:Dalphabeta}
    \mathcal{D}_{\alpha\beta}
        =\text{Tr}\left(\sqrt{\rho_{\beta}^{1/2}
            \rho_{\alpha}\rho_{\beta}^{1/2}}
                \right)
        =\left|\left\langle
            \alpha|\beta\right\rangle
                \right|.
\end{equation}
However, they wish to do this without sharing any more information about their qubit state than strictly necessary. Here we describe a protocol that accomplishes this in the so-called `honest-but-curious' setting -- i.e., we merely seek to secure data from curious prying eyes, but we assume that Alice and Bob are honest at carrying out their tasks and therefore make no attempt at \emph{verifying} computation to guard against erroneous results.

    A simple solution to learning $\mathcal{D}_{\alpha\beta}$ for pure states is the comparator circuit shown in Fig.~\ref{fig:Comparator}. It is easy to show (see Appendix) that under this circuit, the projector $\Pi_{1,1}=\ket1\bra1\otimes\ket1\bra1$
    has an expectation value that directly yields infidelity $\left\langle \Pi_{1,1}\right\rangle =\frac{1}{2}\left(1-\mathcal{D}_{\alpha\beta}^{2}\right)$. If $\rho_{\alpha}$ or $\rho_{\beta}$ is mixed, we must replace $\mathcal{D}_{\alpha\beta}^{2}$ in the last equation with $\mathcal{D}_{\alpha\beta}^{\left(2\right)}=\text{Tr}\left(\rho_{\beta}^{1/2}\rho_{\alpha}\rho_{\beta}^{1/2}\right)$. Note, for consistency, that this reduces back to $\left|\left\langle \alpha|\beta\right\rangle \right|^{2}=\mathcal{D}_{\alpha\beta}^{2}$ for pure states. We build our protocol on this simple comparator circuit with these slight additions.\figComparatorBasic
    
As prescribed by our QFHE scheme, Alice encrypts all $n$ copies of her qubit by preparing $\sigma_{\alpha}^{\left(1\right)}\otimes...\otimes\sigma_{\alpha}^{\left(n\right)}=Z^{\vec{a}}X^{\vec{b}}\rho_{\alpha}^{\otimes n}X^{\vec{b}}Z^{\vec{a}}$ and sends them all at once to Bob along with classically homomorphically encrypted keys $\text{Enc}\left(\vec{a}\right)$ and $\text{Enc}\left(\vec{b}\right)$. Here, $\vec{a},\vec{b}\in\left\{ 0,1\right\} ^{\otimes n}$ and $X^{\vec{a}}$ is to be interpreted as $X^{a_{1}}\otimes\cdots\otimes X^{a_{n}}$.
Bob in turn performs the comparator circuit between each of Alice's qubit $\sigma_{\alpha}^{\left(i\right)}$ and his own $\rho_{\beta}$. Figure~\ref{fig:Secure2PartyComparator} illustrates this. After measuring the $i$-th pair of qubits, Bob homomorphically adds
classical outcomes $k_{1}^{\left(i\right)}$ and $k_{2}^{\left(i\right)}$ to Alice's encrypted keys to obtain $\text{Enc}\left(a_{i}+k_{1}^{\left(i\right)}\right)$ and $\text{Enc}\left(b_{i}+k_{2}^{\left(i\right)}\right)$. Referring to key transformation rules in Table~\ref{tab:CliffordKeyTransform}, note that once Alice decrypts these and computes $\frac{1}{n}\sum_{i}\left\{ \left(a_{i}+k_{1}^{\left(i\right)}\right)\times\left(b_{i}+k_{2}^{\left(i\right)}\right)\right\} $, she correctly obtains $\left\langle \Pi_{1,1}\right\rangle $. Addition \& multiplication in the summands are modulo 2, whereas the top-level summation is on the full set of integers.\figComparatorSecure
    
    An essential step in our protocol is for Bob to scramble the order of $\text{Enc}\left(a_{i}+k_{1}^{\left(i\right)}\right)$ and $\text{Enc}\left(b_{i}+k_{2}^{\left(i\right)}\right)$ before returning to Alice the scrambled classical ciphertexts $\text{Enc}\left(a_{s\left(i\right)}+k_{1}^{\left(s\left(i\right)\right)}\right)$ and $\text{Enc}\left(b_{s\left(i\right)}+k_{2}^{\left(s\left(i\right)\right)}\right)$ where $s\left(i\right)$ is a random permutation on the set $\left\{ 1,...,n\right\} $. This is important in order to ensure the security of Bob's qubit which, unlike Alice's qubit, is \emph{not} encrypted. Absent this scrambling, Alice can prepare and keep track of the inner product between Bob's qubit and a variety of states of her choosing, thereby effectively doing tomography on Bob's state. This completes the description of our protocol.

\subsection{\label{2PSCexp}Experimental Demonstration}
    \figExperimentTPSC We implement this protocol by using a subset of our full setup, with minimal modifications. Figure~\ref{fig:ExperimentalComparator} illustrates this. Computer-controlled waveplates representing Alice and Bob were programmed to randomly select and prepare qubits from a predefined set of logical/plaintext states (determined by range and precision of motion of our motorised actuators). In Alice's case, these waveplate settings take into account randomly generated one-time-pads. The infidelity between Alice's and Bob's states is then measured as described above and plotted against its actual value in Fig.~\ref{fig:2PSC-correct}.
As we have done in previous sections, we also ran the protocol for the same set of logical input states but with an intentionally erroneous decryption key.
The result is shown in Fig.~\ref{fig:2PSC-wrong}. Every point in each of these plots contains $\sim960$ photons or qubits.

Observe in Fig.~\ref{fig:2PSC-correct} that in the case of correct decryption $2\left\langle \Pi_{1,1}\right\rangle $ shows good agreement with theoretically expected values (dashed line), indicating the protocol indeed allows Alice retrieve the infidelity between her state and Bob's. By comparison, in Fig.~\ref{fig:2PSC-wrong} where decryption is done incorrectly, $2\left\langle \Pi_{1,1}\right\rangle$ hovers near a constant $0.5$ (i.e., the infidelity w.r.t. the maximally mixed state), suggesting that anyone without the secret key gains no information about that infidelity.\figTPSCdataGood \figTPSCdataBad

\section{\label{Conc}Discussion and Conclusion}
    In this work we constructed, implemented, and demonstrated a fully homomorphic encryption scheme for universal gate-based quantum computers first proposed in~\cite{StaceyQFHE} and extended in \cite{SpeelmanCompactQFHE}. With this scheme, any party in possession of encrypted qubits may evaluate a quantum circuit of their choice. This is accomplished with the aid of ancillas and classical bits prepared at the time of encryption and transmitted along with the ciphertext. Multiple use of a communication channel, quantum or classical, is not required. Explicit knowledge of the circuit(s) evaluated is not necessary for correct decryption.
    Furthermore, we make no concessions on security apart from assumptions that underlie the classical homomorphic cryptosystem that we use to construct our scheme. Previously demonstrated schemes compromise on one or more of these attributes.

    We also solve the simple task of computing the inner-product of two qubit states securely; that is, without allowing either party to tomographically characterise the other's qubit. Our encryption scheme provides for an elegant solution to this task, which is otherwise difficult to accomplish.
    
    While our scheme is secure in theory, we note that our optical implementation relies on post-selections that do not always succeed.
This, along with practical experimental losses, implies that a significant number of photons that carry the same qubit state will fail to register on our detectors.
Keys that in theory ought to be one-time-pads may no longer actually be used exactly once in practice, and a would-be attacker could siphon off these otherwise `lost' photons
    to gain information about plaintexts. Nevertheless, we stress that these are technical shortcomings that can be remedied in a production setting, for example by greatly increasing key-switching rate (impractical in our setup) or simply using a different physical platform and/or gate design entirely.

\bibliography{bib}

\section{\label{AppendixA}Appendix: Infidelity measure from a simple comparator circuit}
Suppose $\rho_{\alpha}=\ket{\alpha}\left<\alpha\right|$ and
   $\rho_{\beta}=\ket{\beta}\left<\beta\right|$. Now write $\ket{\alpha}$
    and $\ket{\beta}$ in the computational basis:
    \begin{align*}
    \ket{\alpha} & =c_{0}\left|0\right>+c_{1}\left|1\right>\\
    \ket{\beta} & =d_{0}\left|0\right>+d_{1}\left|1\right>
    \end{align*}
    where $c_{i},d_{i}\in\mathbb{C}$ and $\sum_{i}\left|c_{i}\right|^{2}=\sum_{i}\left|d_{i}\right|^{2}=1$.
    We can therefore rewrite:
    \begin{align*}
    \mathcal{D}_{\alpha\beta}^{2} & =\left|\left\langle \alpha|\beta\right\rangle \right|^{2}\\
     & =\left|c_{0}^{*}d_{0}+c_{1}^{*}d_{1}\right|^{2}\\
     & =\left|c_{0}d_{0}\right|^{2}+\left|c_{1}d_{1}\right|^{2}+2\text{Re}\left(c_{0}c_{1}^{*}d_{0}^{*}d_{1}\right)\\
     & =1-\left|c_{0}d_{1}\right|^{2}-\left|c_{1}d_{0}\right|^{2}+2\text{Re}\left(c_{0}c_{1}^{*}d_{0}^{*}d_{1}\right)
    \end{align*}
    Now consider the circuit in Figure \ref{fig:Comparator}. The two-qubit
    state that results is:
    \begin{align*}
    \left|\alpha,\beta\right>\to & \frac{1}{\sqrt{2}}\Big\{\left(c_{0}d_{0}+c_{1}d_{1}\right)\left|0,0\right>+\left(c_{0}d_{0}-c_{1}d_{1}\right)\left|1,0\right>\\
     & +\left(c_{0}d_{1}+c_{1}d_{0}\right)\left|0,1\right>+\left(c_{0}d_{1}-c_{1}d_{0}\right)\left|1,1\right>\Big\}
    \end{align*}
    which immediately yields:
    \begin{align*}
    \left\langle \Pi_{1,1}\right\rangle  & =\frac{1}{2}\left|c_{0}d_{1}-c_{1}d_{0}\right|^{2}\\
     & =\frac{1}{2}\left\{ \left|c_{0}d_{1}\right|^{2}+\left|c_{1}d_{0}\right|^{2}-2\text{Re}\left(c_{0}c_{1}^{*}d_{0}^{*}d_{1}\right)\right\} \\
     & =\frac{1}{2}\left(1-\mathcal{D}_{\alpha\beta}^{2}\right)
    \end{align*}
    Note that if $\rho_{\alpha}$ and/or $\rho_{\beta}$ is/are mixed
    states, cross-terms arise in squaring $\mathcal{D}_{\alpha\beta}$.
    In that case, this last equation holds if we replace $\mathcal{D}_{\alpha\beta}^{2}$
    with:
    \[
    \mathcal{D}_{\alpha\beta}^{\left(2\right)}=\text{Tr}\left(\rho_{\beta}^{1/2}\rho_{\alpha}\rho_{\beta}^{1/2}\right)
    \]

\end{document}